\documentclass[twocolumn,pra,amssymb,showpacs]{revtex4-1}
\usepackage{graphicx}
\usepackage{amsmath}
\usepackage{subfigure}

\newcommand{\be}{\begin{equation}}
\newcommand{\ee}{\end{equation}}
\newcommand{\ben}{\begin{eqnarray}}
\newcommand{\een}{\end{eqnarray}}

\begin{document}

\title{Locality and Classicality: role of entropic inequalities}

\author{J. Batle$^1$, Mahmoud Abdel-Aty$^{2,3}$, C. H. Raymond Ooi$^4$, 
S. Abdalla$^5$ and Y. Al-hedeethi$^5$}
\email{E-mail address (JB): jbv276@uib.es}
\affiliation{
$^1$Departament de F\'{\i}sica, Universitat de les Illes Balears, 07122 Palma de Mallorca, Balearic Islands, Europe \\
$^2$Department of Mathematics, Faculty of Science, Sohag University, Sohag, Egypt\\
$^3$University of Science and Technology at Zewail City, 12588 Giza, Egypt\\
$^4$Department of Physics, University of Malaya, 50603 Kuala Lumpur, Malaysia\\
$^5$Department of Physics, Faculty of Science, King Abdulaziz University Jeddah, P.O. Box 80203, Jeddah 21589, Saudi Arabia\\\\}

\date{\today}

\begin{abstract}

The use of the so-called entropic inequalities is revisited in the light of new quantum correlation measures, specially 
nonlocality. We introduce the concept of {\it classicality} as the non-violation of these classical inequalities by quantum 
states of several multiqubit systems and compare it with the non-violation of Bell inequalities, that is, {\it locality}. 
We explore --numerically and analytically-- the relationship between several other quantum measures and discover the 
deep connection existing between them. The results are surprising due to the fact that these measures are very different 
in their nature and application. The cases for $n=2,3,4$ qubits and a generalization to systems with arbitrary number of 
qubits are studied here when discriminated according to their degree of mixture. 

\end{abstract}



\maketitle

\section{Introduction}

Entanglement is perhaps one of the most fundamental and non-classical features exhibited by quantum
systems \cite{LPS98}, that lies at the basis of some of the most important
processes studied by quantum information theory \cite{Galindo,NC00,LPS98,WC97,W98}
such as quantum cryptographic key distribution \cite{E91}, quantum
teleportation \cite{BBCJPW93}, superdense coding \cite{BW93}, and quantum
computation \cite{EJ96,BDMT98}. It is plain from the fact that entanglement
is an essential feature for quantum computation or secure quantum
communication, that one has to be able to develop some procedures (physical
or purely mathematical in origin) so as to ascertain whether the state $\rho$
representing the physical system under consideration is appropriate for
developing a given non-classical task. 

That is, it is essential to discriminate the states that contain
classical correlations only. Historically, the violation of Bell's inequalities
have become equivalent to non-locality or, in our this context,
to entanglement. For every pure entangled state there is a Bell inequality
that is violated and, in consequence, from a historic viewpoint, the first
separability criterion is that of Bell (see \cite{T02} and references therein).
It is not known, however, whether in the case of many entangled mixed states,
violations exist: some states, after ``distillation" of entanglement (this is
done by performing local operations and classical operations (LOCC), that is,
operations performed on each side independently) eventually violate the
inequalities, but some others do not.

The first to point out that an entangled state did not imply violation of
Bell-type inequalities (that is, they admit a local hidden variable model)
was Werner \cite{Werner89}, providing himself with a family of mixed states
(the {\it Werner} $\rho_{W}$) that do no violate the aforementioned
inequalities. Werner also provided the current mathematical definition for
separable states: a state of a composite quantum system
constituted by the two subsystems $A$ and $B$ is called ``entangled" if it can
not be represented as a convex linear combination of product states. In other
words, the density matrix $\rho_{AB} \in {\cal H}_A \otimes {\cal H}_B$
represents an entangled state if it cannot be expressed as

\begin{equation} \label{sepa}
\rho_{AB} \, = \, \sum_k \, p_k \, \rho_{A}^{(k)}
\otimes \rho_{B}^{(k)},
\end{equation}

\noindent with $0\le p_k \le 1$ and $\sum_k p_k =1$. On the contrary, states of
the form (\ref{sepa}) are called separable. The above definition is physically
meaningful because entangled states (unlike separable states) cannot be
prepared locally by acting on each subsystem individually \cite{P93}
(LOCC operations). 
In practice, to find whether an arbitrary state $\rho$ can be expressed like (\ref{sepa})  is an impossible task because
there are infinitely
many ways of decomposing a state $\rho$ (for instance, the pure states constituting
the alluded mixture need not be orthogonal, what makes it even more arduous).
Physically, it means that
the state can be prepared in many ways. 

The separability question had quite interesting  echoes in information theory
and its associate information measures or entropies. When one deals with a
classical composite system described by a suitable probability distribution
defined over the concomitant phase space, the entropy of any of its subsystems
is always equal or smaller than the entropy characterizing the whole system.
This is also the case for separable states of a composite quantum
system \cite{NK01,VW02}. In contrast, a subsystem of a quantum system described
by an entangled state may have an entropy greater than the entropy of
 the whole system. Indeed, the von Neumann entropy of either
 of the subsystems of a bipartite quantum system described (as a whole)
 by a pure state provides a natural measure of the amount of entanglement
 of such state. Thus, a pure state (which has vanishing entropy)
 is entangled if and only if its subsystems have an entropy
 larger than the one associated with the system as a whole. 
 The previous facts constitute the ``classical entropic inequalities" which, in terms of 
 conditional entropies, read as

  \ben 
  S(A|B) &\ge & 0, \cr
  S(B|A) &\ge & 0,
  \een

  \noindent accomplished by all separable states \cite{Horo99} ($S$ stands for the usual von Neumann entropy). 
  The early motivation for the studies reported in
  \cite{VW02,HHH96,HH96,CA97,V99,TLB01,TLP01,A02} was
  the development of practical separability criteria for density matrices.
  The discovery by Peres of the partial transpose criteria, which for
  two-qubits and qubit-qutrit systems turned out to be both necessary
  and sufficient, rendered that original motivation somewhat outmoded.
However, their study provide a more physical insight into the issue of quantum
separability. In the present contribution, we shall revisit their role as compared to 
other quantum measures and, specially, nonlocality.
\newline

Ever since the introduction of the entropic inequalities, several measures have recently appeared 
in the literature that are not directly related to 
entanglement, but that in some cases --specially when dealing with systems of qubits greater 
that two-- provide a satisfactory approximate answer, like the maximum violation 
of a Bell inequality, that is, nonlocality. Local Variable Models (LVM) cannot exhibit arbitrary correlations. 
Mathematically, the conditions these correlations
must obey can always be written as inequalities --the Bell inequalities-- satisfied for the joint
probabilities of outcomes. We say that a quantum state $\rho$ is nonlocal if and only if there
are measurements on $\rho$ that produce a correlation that violates a Bell inequality. 
Later
work by Zurek and Ollivier \cite{olli} established that not even
entanglement captures all aspects of quantum correlations. These
authors introduced an information-theoretical measure, quantum
discord, that corresponds to a new facet of the ``quantumness"
that arises even for non-entangled states. Indeed, it turned out
that the vast majority of quantum states exhibit a finite amount
of quantum discord. Besides its intrinsic
conceptual interest, the study of quantum discord may also have
technological implications: examples of improved quantum computing
tasks that take advantage of quantum correlations but do not rely
on entanglement have been reported [see for instance, among a
quite extensive references-list \cite{geom,olli,ferraro,dattaprl,luo}. 
Actually, in some cases entangled states are useful to solve a problem if and only if they
violate a Bell inequality \cite{comcomplex}. Moreover, there are
important instances of non-classical information tasks that are
based directly upon non-locality, with no explicit reference to
the quantum mechanical formalism or to the associated concept of
entanglement \cite{device}. A recent work studying how entanglement can be estimated from a Bell inequality violation also sheds new light on the use of Bell inequalities \cite{new}. 
Although we are going to use the maximization of the Clauser-Horne-Shimony-Holt inequality \cite{CHSH} for the case of two qubits, we have to bear in mind that other inequalities can be used as well, 
such as the I3322 inequality (the simplest bipartite 
two-outcome Bell inequality beyond CHSH) \cite{I3322}.\newline

It is the aim of the present work to focus on the role played by these (classical) 
entropic inequalities when compared to other measures introduced in order to measure 
how ``quantum'' a state is and, specially, the nonlocality measure. This contribution is divided as follows: in Section II we discuss 
the connections between entropic inequalities and other measures. We will provide some theorems that highlight the intimate 
(and unexpected) connection with the violation of a Bell inequality when studying the whole set of mixed states of two qubits, and when they are 
discriminated according to two different measures of the mixedness of the state $\rho$. A similar study for higher systems will be conducted as well, specifically for 
three and four qubits. A general insight into systems of arbitrary number of qubits will also be presented. Finally, we shall draw some conclusions in Section III. Two appendices appear related to the description of the quantum measures used and the methods employed, as well as how to generate random two qubit states according to some given measure, and how to obtain a particular subset of them endowed with a given degree of mixture.

\section{Results}

\subsection{Two qubits}

The current status of the type of quantal correlations present in a given state of a system has been recently extended by the notion of quantum discord. Thus, in addition to entanglement, other measures for quantum discord (the original one and a complementary one, the so-called geometric quantum discord), or the maximal violation of Bell inequalities, are used as detectors of the ``quantumness" or ``nonlocality" of a state. In this scenario, the use of classical entropic inequalities has not attracted much attention for stronger criteria are more useful. However, it is of our interest to compare how good the entropic inequalities are when compared to the previous set of measures, that is, entanglement, quantum discord, geometric discord and maximal violation of a Bell inequality. Entropic inequalities constitute a clear physical and information-theoretical criterion for discussing the ``classicality" of a quantum state. No other criteria approaches the entropic ones because they lack the required physical intuition, which is applied in thermodynamic systems: {\it the whole is greater than the sum of its parts}.

For two qubits, the entropic inequalities read

\begin{eqnarray} \label{entropic2}
S(\rho_{AB})-S(\rho_A) &\geq& 0\\
S(\rho_{AB})-S(\rho_B) &\geq& 0,
\end{eqnarray}

\noindent where $S$ is the usual von Neumann entropy. In terms of eigenvalues of the state and reduced states, we have the more convenient expressions

\begin{eqnarray}
\lambda_1^{\lambda_1} \lambda_2^{\lambda_2} \lambda_3^{\lambda_3} \lambda_4^{\lambda_4}
&\leq& \alpha^{\alpha} (1-\alpha)^{1-\alpha} \\
\lambda_1^{\lambda_1} \lambda_2^{\lambda_2} \lambda_3^{\lambda_3} \lambda_4^{\lambda_4}
&\leq& \beta^{\beta} (1-\beta)^{1-\beta},
\end{eqnarray}

\noindent with $\{\alpha,1-\alpha,\beta,1-\beta\}$ being the eigenvalues for the reduced states of each party, and $\lambda_i$ are the eigenvalues of $\rho_{AB}$ which, without loss of generality, can verify the condition $\lambda_1 > \lambda_2 > \lambda_3 > \lambda_4$.

It seems plausible then to assume that the magnitude of the violation (having a negative sign) of (\ref{entropic2}) must possess some kind of positive correlation with the previous aforementioned measures. Thus, an exploration of the whole set of mixed states of two qubits is mandatory to ascertain how good two measures coincide with each other.

The concomitant figures Fig. 1 a), Fig. 1 b), Fig. 1 c) and Fig. 1 d) depict, respectively, the concurrence, the quantum discord, the geometrical discord and the maximum violation of the CHSH Bell inequality for a given amount of the smallest conditional entropy of either two subsystems of two qubits. In each case, the solid line depicts a typical Werner state for comparison. In all cases, we can appreciate a positive tendency for all quantities. However, in the case of discords, since the set of states with zero discord has null measure, they do not appear in the current exploration, and therefore there is no minimum quantity reached. As explained in Appendix II, states $\rho$ are generated according to the product measure $\nu = \mu \times {\cal L}_{N-1}$, where $\mu$ is the Haar measure on the group of unitary matrices ${\cal U}(N)$, along with the standard normalized Lebesgue measure ${\cal L}_{N-1}$ on ${\cal R}^{N-1}$.

\begin{figure}[htbp]
\begin{center}
\includegraphics[width=8.8cm]{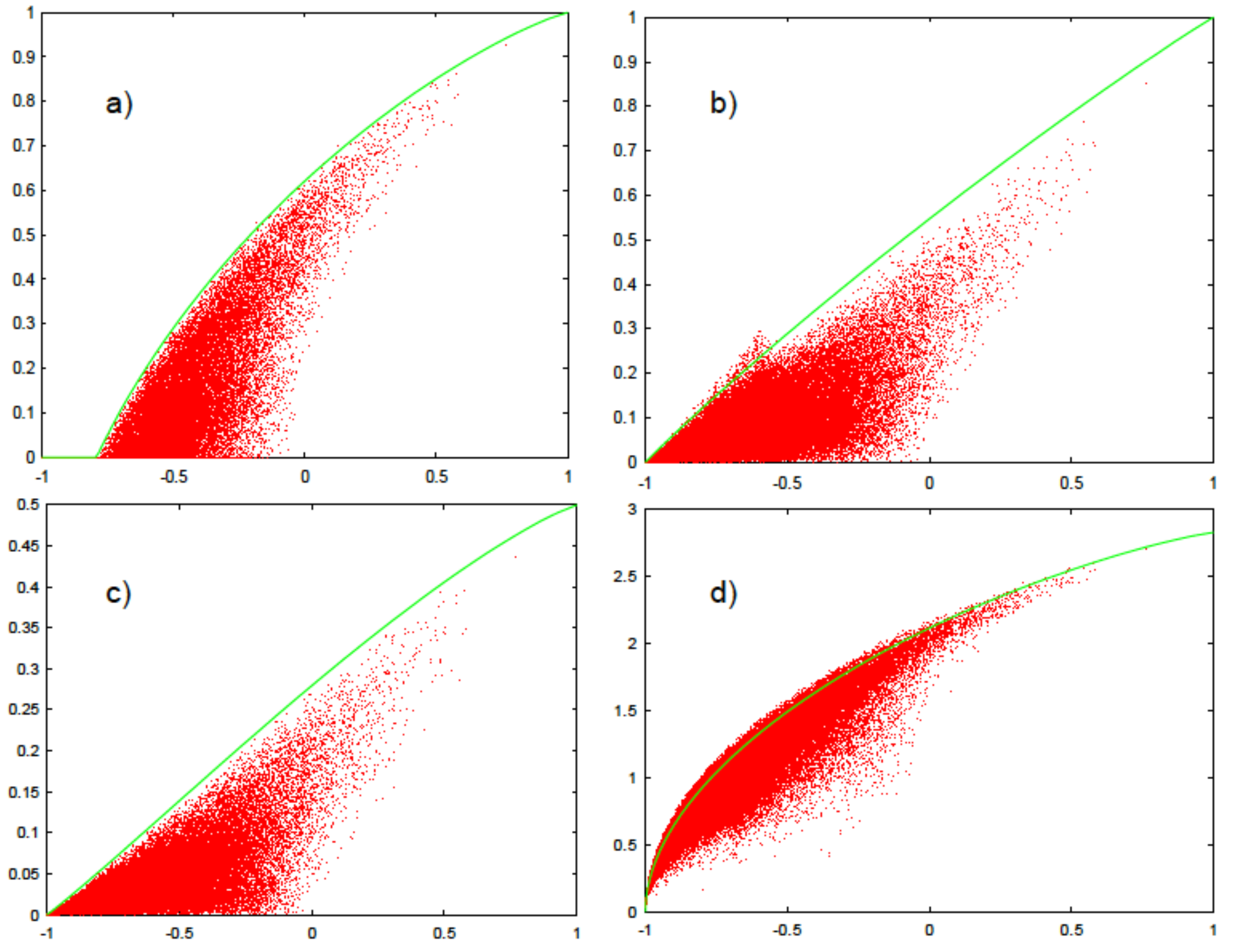}
\caption{(Color online) Plot of the dependency of a) concurrence, b) quantum discord, c) geometric discord and c) maximum violation of CHSH Bell inequality vs. the minimum of the negative quantities in (\ref{entropic2}) in units of $\ln 2$, for a sample of $10^6$ random states. In all cases, we appreciate a strong correlation. The solid (green) line depicts the case for Werner states. See text for details.}
\label{fig1}
\end{center}
\end{figure}

What is depicted in Fig. 1 are several quantum correlation measures versus the minimum of both two quantities appearing in the definition of conditional entropies. From the figure, we can see that the areas with low values of quantum correlations are strongly populated with states that violate the entropic inequalities (negative horizontal axis). Although considerably less, one can find states (very few) not violating any entropic inequality (positive horizontal axis) with high values of quantum correlations, which proves that entropic criteria fail considerably. The functional dependency between violation of entropic inequalities and existence of non-zero quantum correlations must be interpreted as follows: states violating the entropic criteria lie very close to zero or small quantum correlation values. Staying within the negative horizontal axis, as we move to the right, we tend to find less population of states with higher and higher values of quantal correlations. Therefore, the correlation between entropic measures (their violation) and quantum measures should be better understood if we restrict ourselves to the negative horizontal axis.

The numerical survey over all states (a sample of a million mixed states in our case) displays very interesting results when i) the whole set of states is considered, and when ii) states are discriminated according to their degree of mixture $R=1/Tr(\rho^2)$. We could have used the so-called purity 
$Tr(\rho^2)$ instead as a degree of mixture. However, one being the inverse of the other, they provide the same information regarding the degree of mixture $R$ of the state $\rho$. When we consider the probability that the entropic criteria and any other measure providing the same answer, we arrive at these corresponding results:

\begin{itemize}

\item entropic inequalities and entanglement: there is a probability of $64.3\%$ that an arbitrary state either violates (fulfills) the entropic inequalities with non-zero (null) entanglement.

\item entropic inequalities and the CHSH Bell inequality: there is a probability of $99.4 \%$ that an arbitrary state either violates (fulfills) the entropic inequalities with maximal CHSH violation less than --that is, local-- (greater than --that is, nonlocal--) two.

\end{itemize}

\noindent In addition, it is seen that both entanglement and nonlocality have a concomitant
probability of $64.6\%$ of coincidence.

\noindent When we use a particular family of states, the Bell diagonal states, the situation becomes the following:

\begin{itemize}

\item entropic inequalities and entanglement: there is a probability of $83.6\%$ that an arbitrary state either violates (fulfills) the entropic inequalities with non-zero (null) entanglement.

\item entropic inequalities and the CHSH Bell inequality: there is a probability of $99.4\%$ that an arbitrary state either violates (fulfills) the entropic inequalities with maximal CHSH violation less than --that is, local-- (greater than --that is, nonlocal--) two.

\end{itemize}

\noindent Entanglement and nonlocality have a concomitant probability of $84.2\%$ of coincidence. As it can be appreciated, there is a considerable, almost perfect correlation between entropic inequalities and nonlocality, whereas the coincidence is lower when comparing with entanglement.

When states are generated with a given value of $R$, the concomitant probabilities of coincidence
display a very surprising behavior. In Fig. 2, the probability of both entropic criterion and entanglement leading to the same answer (lower curve) and the same thing for entropic and nonlocality (upper curve) are depicted, respectively. It is apparent from our calculations that all states with degree of mixture $R$ between 2 and 3 do not violate any entropic inequality, and the same thing happens for nonlocality: all states are local as well, that is, their maximal CHSH Bell quantity is always less than or equal to 2. Thus, both entropic and Bell inequalities provides the same answer in the aforementioned region. Therefore, the probability of coincidence is exactly 1.

\begin{figure}[htbp]
\begin{center}
\includegraphics[width=8.8cm]{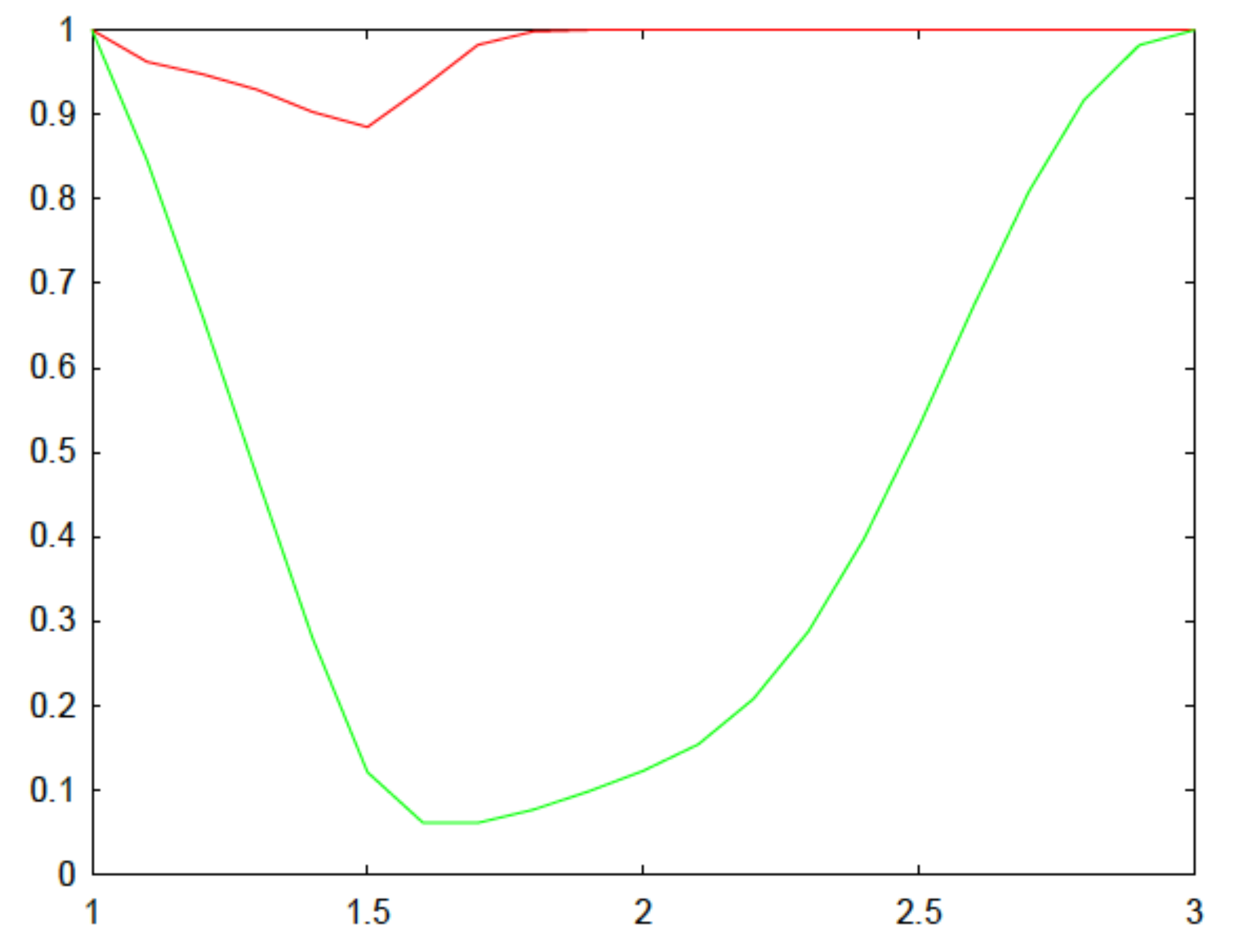}
\caption{(Color online) Probability of coincidence between concurrence and entropic inequalities (lower curve), and coincidence between CHSH Bell inequalities and entropic inequalities (upper curve) vs. $R$. 
In the latter case, small discrepancies appear nearby pure states ($R=1$). Every point has been obtained generating $10^6$ random states with the same mixture $R$ (see Appendix II). See text for details.}
\label{fig2}
\end{center}
\end{figure}

Now we are going to summarize and to prove the previous facts with respect to all states with a given degree of mixture, either given by $R$ or the maximum eigenvalue $\lambda_m$ of a state $\rho$, in the following theorems:
\newline

{\it Proposition 1}: If the participation ratio $R=1/Tr(\rho^2)$ is given to be $R \in [2,4]$, then the maximum eigenvalue $\lambda_m$ of $\rho$ is bounded inside $\lambda_m \in [\frac{1}{4},1/\sqrt{2} \approx 0.7)$.

{\it Proof:} The condition $R \geq 2$ is equivalent to $\lambda_1^2+\lambda_2^2+\lambda_3^2+\lambda_4^2 \leq 1/2$. Now, the minimum value for $\lambda_1$ is clearly 1/4 because no eigenvalue is greater that their arithmetic mean. The maximum is obtained as follows: From the squares of the eigenvalues, it is plain 
that $\lambda_1^2<1/2$, or $\lambda_1<0.7$. Thus, $\lambda_1 \in [\frac{1}{4},0.7)$ $\blacksquare$ 
\newline

{\it Proposition 2}: If the maximum eigenvalue of a state $\rho$ is given to be $\lambda_m \leq 1/2$, then we have $R \in [2,4]$.

{\it Proof:} If $\lambda_1 \leq 1/2$, then the same condition applies to all the rest 
$\lambda_i \leq 1/2$. Therefore, we can define the quantities $\lambda_i (1/2-\lambda_i)$ to be obviously positive. When developing them, we have $\frac{1}{2}\lambda_i \geq \lambda_i^2$. By adding the whole set of inequalities $i=1..4$, we encounter that $\lambda_1^2+\lambda_2^2+\lambda_3^2+\lambda_4^2 \leq 1/2$, which is tantamount as having $R \geq 2$ $\blacksquare$ \newline

\noindent With these two propositions, we are about to prove one of the major results of the present work for two qubit states.\newline

{\it Theorem 1 ($R$-language)} For all mixed states $\rho$ fulfilling $R \in [2,4]$, these two statements are equivalent:

\begin{itemize}
\item There is no violation of the entropic inequalities
\item No Bell inequality is violated (specifically, the CHSH inequality)
\end{itemize}

{\it Proof:} Let us proof the first statement. The non-violation of entropic inequalities requires $S(\rho_{AB})-S(\rho_A) \geq 0$ and $S(\rho_{AB})-S(\rho_B) \geq 0$. Suppose that
$\lambda_1>\lambda_2>\lambda_3>\lambda_4$ is the set of eigenvalues of an arbitrary state of two qubits. Formally, we have that $-\lambda_1\ln \lambda_1 -\lambda_2\ln \lambda_2 -\lambda_3\ln \lambda_3
-\lambda_4\ln \lambda_4 - (-\alpha\ln \alpha - (1-\alpha)\ln (1-\alpha)) \geq 0$, where $\alpha$ and $1-\alpha$ are the eigenvalues of the reduced density matrices ($\alpha \in [0,1]$). The same expression can be written more compactly as

\begin{equation} \label{bloc}
\lambda_1^{\lambda_1} \lambda_2^{\lambda_2} \lambda_3^{\lambda_3} \lambda_4^{\lambda_4} \leq \alpha^{\alpha} (1-\alpha)^{(1-\alpha)}
\end{equation}

\noindent $\alpha^{\alpha} (1-\alpha)^{(1-\alpha)}$ is a monotonically decreasing
function of $\alpha$, ranging from 1 to 1/2 for $\alpha \in [0,1]$. Since, according to Proposition 1,
we have that the maximum eigenvalue $\lambda_1 \leq 0.7$, then all other eigenvalues $\lambda_i$ must also be less than 0.7. Plugging the aforementioned values into (\ref{bloc}), we have

\begin{equation} \label{07}
0.7^{0.7} \cdot 0.7^{0.7} \cdot 0.7^{0.7} \cdot 0.7^{0.7} <
0.37 < \alpha^{\alpha} (1-\alpha)^{(1-\alpha)} < 1.
\end{equation}

\noindent That is, no violation of the entropic inequalities occurs $\blacksquare$
\newline

Let us prove the second statement. It will be suitable to have the arbitrary state of two qubits written in the Bloch representation

\begin{equation} 
4\rho=   \mathcal{I} \otimes \mathcal{I} +
\sum_{u=1}^{3} x_u \sigma_u \otimes \mathcal{I} + \sum_{u=1}^{3}
y_u \mathcal{I} \otimes \sigma_i +\sum_{u,v=1}^{3} T_{uv}
\sigma_u \otimes \sigma_v,
\end{equation}

\noindent with $x_u=Tr(\rho(\sigma_u \otimes \mathcal{I}))$, $y_u=Tr(\rho (\mathcal{I}
\otimes \sigma_u))$, and $T_{uv}=Tr(\rho (\sigma_u \otimes \sigma_v))$. These parameters form the correlation matrix (\ref{Rmatrix}). Let us multiply the state by itself and take the trace. Then $Tr(\rho^2)=\frac{1}{4} [1+ \sum_i x_i^2 + \sum_i y_i^2 + \sum_{i,j} T_{ij}^2 ]$, which is equal to $1/R$. $R \geq 2$ implies $1/R \leq 1/2$, which combined with the previous quantities gives us

\begin{equation}
\sum_i x_i^2 + \sum_i y_i^2 + \sum_{i,j} T_{ij}^2 \leq 1.
\end{equation}

\noindent The former fact implies that

\begin{equation} \label{bound}
\sum_{i,j} T_{ij}^2 \leq 1 - (\sum_i x_i^2 + \sum_i y_i^2 ),
\end{equation}

\noindent which it has to be a positive quantity. From the work developed by Horodecki {\it et al} in \cite{HoroBell} regarding necessary and sufficient conditions for the violation of the CHSH Bell inequality, we have that the quantity
$M(\rho) \equiv max_{\bf{c},\bf{c'}} \big( ||T\bf{c}||^2 + ||T\bf{c'}||^2 $, where $T$ is the matrix defined by $T_{ij}$ ($\bf{c,c'}$ are suitable unit vectors), has to be less than 1 in order not to violate the CHSH Bell inequality. Since the bound $M(\rho) \leq \sum_{i,j} T_{ij}^2$ holds, it is plain from (\ref{bound})
that $M(\rho) \leq 1 - (\sum_i x_i^2 + \sum_i y_i^2) < 1$, which finishes the proof $\blacksquare$

As we can appreciate, the previous result is similar to the one relating unentangled states for all $R \geq 3$, although in our case we deal with nonlocality. Thus, if a state is mixed enough, it will not possess any nonlocality from $R=2$ onward, yet it can be entangled. However, if it is is much more mixed, then all entanglement disappears.

If we measure the degree of mixture by the maximum eigenvalue of the state, then we shall have a similar theorem:\newline

{\it Theorem 2 ($\lambda_m$-language)} For all mixed states $\rho$ fulfilling $lambda_m \leq 1/2$, these two statements are equivalent:

\begin{itemize}
\item There is no violation of the entropic inequalities
\item No Bell inequality is violated (specifically, the CHSH inequality)
\end{itemize}

{\it Proof:} The first sentence can be easily proved in the same fashion that in Theorem 1. Substituting $0.7$ by $0.5$ the bound still holds and that finishes the proof 
$\blacksquare$\newline

The second sentence is proved in a straightforward manner. Owing to Proposition 2, $\lambda_m \leq 1/2$
implies $R \geq 2$, which constitutes the second statement of Theorem 1 $\blacksquare$ \newline

We can summarize the situation in the following way:

\begin{itemize}
\item For $R \in [1,2)$  nonlocality, violation of entropic inequalities and entanglement coexist
\item For $R \in [2,3]$  nonlocality disappears, entropic inequalities are not violated and entanglement is still present
\item For $R \in [3,4]$ neither nonlocality, nor violation of entropic inequalities nor entanglement are present
\end{itemize}

\noindent or use the $\lambda_m$ approach instead: for any state with $\lambda_m \leq 2$ both non violation of entropic inequalities and CHSH Bell inequalities occur. For $\lambda>1/2$, Bell inequalities can be violated (they are in all cases for Bell diagonal states), whereas entanglement is not always present (certainly it is in the Bell diagonal case only). It is worth mentioning that when discriminating the values for the conditional entropies for all states vs. either $R$ or $\lambda_m$, there appear several regions that are bounded by certain states, which will be described somewhere else.

Summing up, we have described a positive correlation between usual quantum measures such as discord, entanglement or nonlocality and the violation of entropic inequalities, quantified by the conditional entropy of each subsystem A and B.
In addition, in the simplest case of two qubits, the degree of coincidence between entropic and Bell inequalities is surprisingly high, which is surprising taking into account the different nature of the magnitudes being represented. This previous fact allow us to shed new light on the classical nature of entropic inequalities for witnessing entanglement and the concomitant relation with nonlocality, given by the violation of the CHSH Bell inequality.

\subsection{Three qubits}

In the case of three qubits, no other measure similar to discord exists. Only entanglement and nonlocality, which is given by the maximum violation of the Mermin inequality, described in Appendix II along with the concomitant optimization techniques. The Mermin, Ardehali, Belinskii and Klyshko (MABK) inequalities (for $n=3$, the Mermin one) are not the only existing Bell inequalities for
$n$ qubits \cite{MABKnew,Ref1,Ref2}, but they constitute a simple generalization of the CHSH one to the multipartite case. Accordingly, it will suffice to use these particular inequalities to
illustrate the basic results of the present work, as far as
non-locality is concerned.

For three qubits, the entropic inequalities read

\begin{eqnarray} \label{entropic3}
S(\rho_{ABC})-S(\rho_A) &\geq& 0\\
S(\rho_{ABC})-S(\rho_B) &\geq& 0\\
S(\rho_{ABC})-S(\rho_C) &\geq& 0,
\end{eqnarray}

\noindent where $S$ is the usual von Neumann entropy $S_1$. When developing into the corresponding states and reduced states, we have the corresponding expression for the spectra of eigenvalues

\begin{eqnarray} \label{entropic3alpha}
\lambda_1^{\lambda_1} \lambda_2^{\lambda_2} \lambda_3^{\lambda_3} \lambda_4^{\lambda_4} \lambda_5^{\lambda_5} \lambda_6^{\lambda_6} \lambda_7^{\lambda_7} \lambda_8^{\lambda_8}
&\leq& \alpha^{\alpha} (1-\alpha)^{1-\alpha} \\
\lambda_1^{\lambda_1} \lambda_2^{\lambda_2} \lambda_3^{\lambda_3} \lambda_4^{\lambda_4} \lambda_5^{\lambda_5} \lambda_6^{\lambda_6} \lambda_7^{\lambda_7} \lambda_8^{\lambda_8}
&\leq& \beta^{\beta} (1-\beta)^{1-\beta} \\
\lambda_1^{\lambda_1} \lambda_2^{\lambda_2} \lambda_3^{\lambda_3} \lambda_4^{\lambda_4} \lambda_5^{\lambda_5} \lambda_6^{\lambda_6} \lambda_7^{\lambda_7} \lambda_8^{\lambda_8}
&\leq& \gamma^{\gamma} (1-\gamma)^{1-\gamma},
\end{eqnarray}

\noindent where we can assume similar conditions ($n=2$ qubits) for the eigenvalues $\{ \lambda_i\}$ of the state $\rho_{ABC}$ and those ones $\{\alpha,1-\alpha,\beta,1-\beta,\gamma,1-\gamma\}$ for the reduced density matrices.

Our numerical survey taken on the whole set of mixed states of three qubits shows that for $R \leq 1.5$ all states violate (\ref{entropic3}). We are going to provide a rough estimate for the boundary as follows. Suppose that we relax the conditions in (\ref{entropic3alpha}) such that all reduced eigenvalues can be reduced to a single expression. In addition, let us remember that the maximum eigenvalue of $\rho_{ABC}$ is given by $\lambda_1$. Then, all other eigenvalues would be
less that $\lambda_1$, which leads us to the following inequality to fulfill the entropic inequalities:

\begin{equation}
\lambda_1^{\lambda_1} \lambda_2^{\lambda_2} \lambda_3^{\lambda_3} \lambda_4^{\lambda_4} \lambda_5^{\lambda_5} \lambda_6^{\lambda_6} \lambda_7^{\lambda_7} \lambda_8^{\lambda_8}
< (\lambda_1^{\lambda_1})^8 \leq \alpha^{\alpha} (1-\alpha)^{1-\alpha} < 1.
\end{equation}

\noindent Let us find $\lambda_1$ such that the inequality is independent of $\alpha$ by setting the equation $(\lambda_1^{\lambda_1})^8 = 1/2$ (recall that $\alpha^{\alpha} (1-\alpha)^{1-\alpha}$ is bounded between 1/2 and 1). The solutions to the previous transcendental equation are 0.90909 and 0.0229, but only the first one is valid because $\lambda_1>1/8$. It can be easily proved that
$\lambda_1=0.90909$ implies a maximum $R \approx 1.2$. Thus, a rough estimate of the minimum $R$ such that all states do not violate (\ref{entropic3}) is found, which is a bit smaller than the numerically observed 1.5.

Not much is know about nonlocality and entanglement in more than two parties. However, some important results have been obtained which are relevant. Following the exact treatment of maximum nonlocality in multiqubit states \cite{JPhysAnostro2011}, it is found that

\begin{equation}\label{MerminR}
Mermin^{\max} (R) \propto \sqrt{\frac{8-R}{8R}},
\end{equation}

\noindent where the constant in (\ref{MerminR}) is obtained by requiring $Mermin^{\max}$ to be equal to 4 for pure states ($R=1$).
One class of states that possess
the previous optimal value is $\rho^{diag}=(1-7x,x,x,x,x,x,x,x)$, which is, interestingly enough, the generalized Werner state for three qubits

\begin{equation}\label{MerminW}
\rho_W^{n=3}= \tilde{x}|GHZ\rangle \langle GHZ| + \frac{1-\tilde{x}}{8} I_8,
\end{equation}

\noindent where $I_8$ is the $8\times8$ identity matrix, and $\tilde{x}=1-8x$. Notice that this was not the case for the two qubit instance.

This interesting feature enables us to discuss the different ranges for $R$ where
to compare nonlocality and presence of genuine tripartite entanglement. On the one hand, from (\ref{MerminR}) we obtain the nonlocality
critical value $R_1=32/11\approxeq 2.9$: {\it no three qubit states possess any nonlocality for participation ratios $R \geq R_1$}. On the other hand,
the special nature of generalized Werner states allow us to compute the separability threshold between entanglement
and separability \cite{WernerThreshold}. From Ref. \cite{WernerThreshold}, the contribution $\tilde{x}$ in (\ref{MerminW}) is such that $x\leq1/5$ involves absence of
entanglement. Translated into $R$-language, it implies a second critical value $R_2=25/4=6.25$:
{\it no three qubit states possess entanglement for participation ratios $R \geq R_2$}.

\begin{figure}[htbp]
\begin{center}
\includegraphics[width=8.8cm]{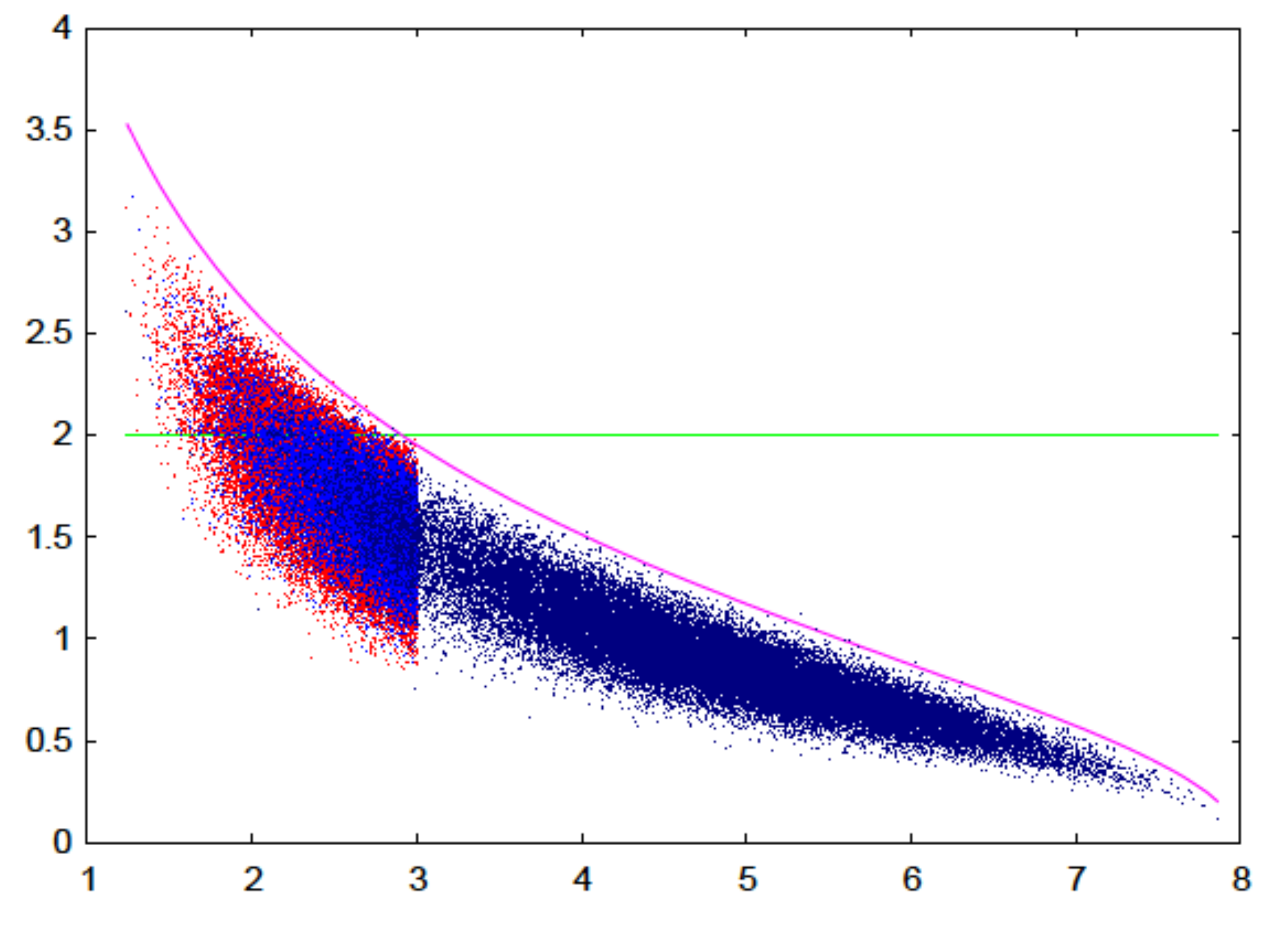}
\caption{(Color online) Plot of the maximum violation of the Mermin inequality for $n=3$ qubits vs. $R$. The region between $R=1$ and $R=3$ has been enhanced so that it can be appreciate that all states beyond $R_1=32/11\approxeq 2.9$ are local. See text for details.}
\label{fig3}
\end{center}
\end{figure}

Therefore, the range of $R$-values splits into three regions: i) between 1 (pure states) and $R_1$, maximum amounts of nonlocality imply the presence of entanglement;
ii) between $R_1$ and $R_2$ we have no violation of the Mermin inequality, yet there exists entanglement; finally, iii) region between $R_2$ and $R=8$
(maximally mixed state) displays absence of both magnitudes. We have to mention that our numerical survey for all state agrees with the nonlocality boundary described previously. In Fig. 3 a sample of states is shown where the maximum violation of the Mermin inequality is depicted vs. the degree of mixture $R$. The solid line corresponds to (\ref{MerminR}). Notice that the Mermin inequality is violated for values greater than 2.

When gathering the results obtained from the entropic inequalities, the whole situation looks quite interesting, as depicted in Fig. 4. The boundaries for entropic inequalities are not strictly defined
analytically, whereas the nonlocality region is exact. The region with zero entanglement is estimated via the only known result for three qubits, that is, the Werner state. Therefore it is likely that $R_2$ could be a bit bigger than the actual (unknown) value. In any case, we can appreciate the great extraordinary coincidence between the non-violation of the classical entropic inequalities and the non-violation of the Mermin Bell inequality, in addition to a hierarchy of criteria entropic-nonlocality-entanglement as the degree of mixture increases. In other words, if classicality is described by fulfilling the entropic inequalities and locality by not violating any Local Variable Model bound, we have found here that both quantities coincide to a very great extend, although being very different in nature. Only for states quite close to pure states most quantum measures coincide.

\begin{figure}[htbp]
\begin{center}
\includegraphics[width=8.8cm]{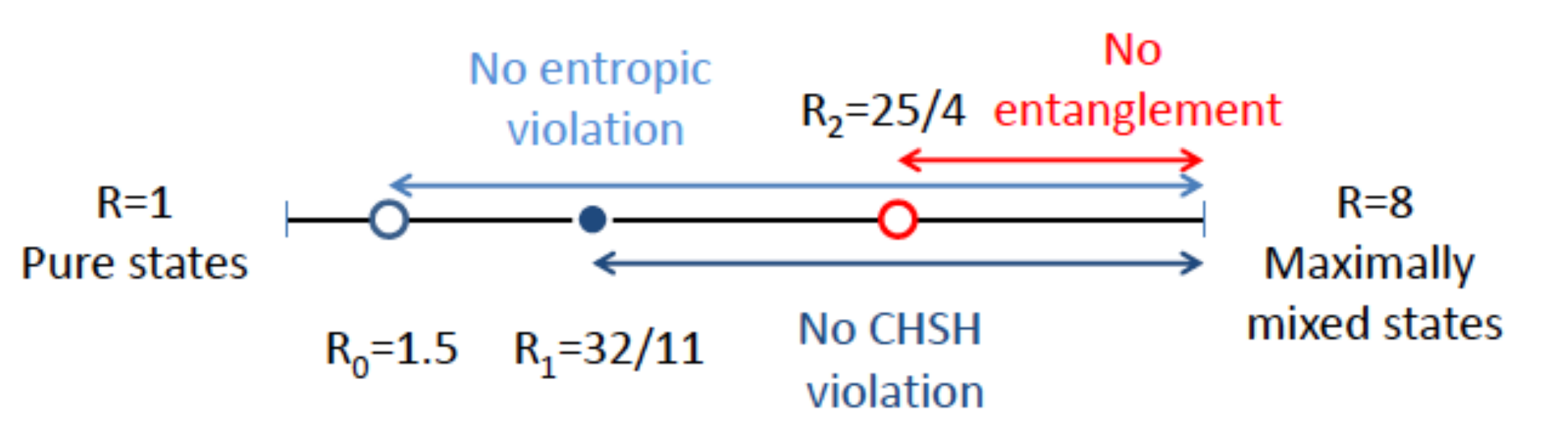}
\caption{(Color online) Plot of the situation regarding locality, reality and entanglement ranges when states of three qubits are discriminated according to their degree of mixture $R$. See text for details.}
\label{fig_line}
\end{center}
\end{figure}

\subsection{Four qubits}

When discussing higher number multiqubit states, our knowledge about quantum correlations is very limited. Alas, no real measure for entanglement exists and only violation of maximum Bell inequalities can be performed. Furthermore, the application of the entropic inequalities becomes quantitatively more involved, since a set of $2^{n-1}-1$ inequalities must be studied, which correspond to all bipartite divisions of the state $\rho$ into subsystems.

However, our numerical exploration shows in Fig. 5 that all states are likely to fulfill the concomitant entropic inequalities for $R \in [3,16]$, whereas all states seem to fulfill the locality requirements for the corresponding MABK Bell inequality for $R \in [4,16]$. What we see again is a great coincidence between locality and classicality, the same hierarchy as in previous cases and a dilatation of the regions where they both appear. It seems plausible to think that the tendency for higher $n$ systems is that the region where locality and classicality coincide extends, which provides a useful tool for ascertaining nonlocality (and entanglement) through the use of a simple, classical entropic criteria. And, indeed, this is the case.

\begin{figure}[htbp]
\begin{center}
\includegraphics[width=8.8cm]{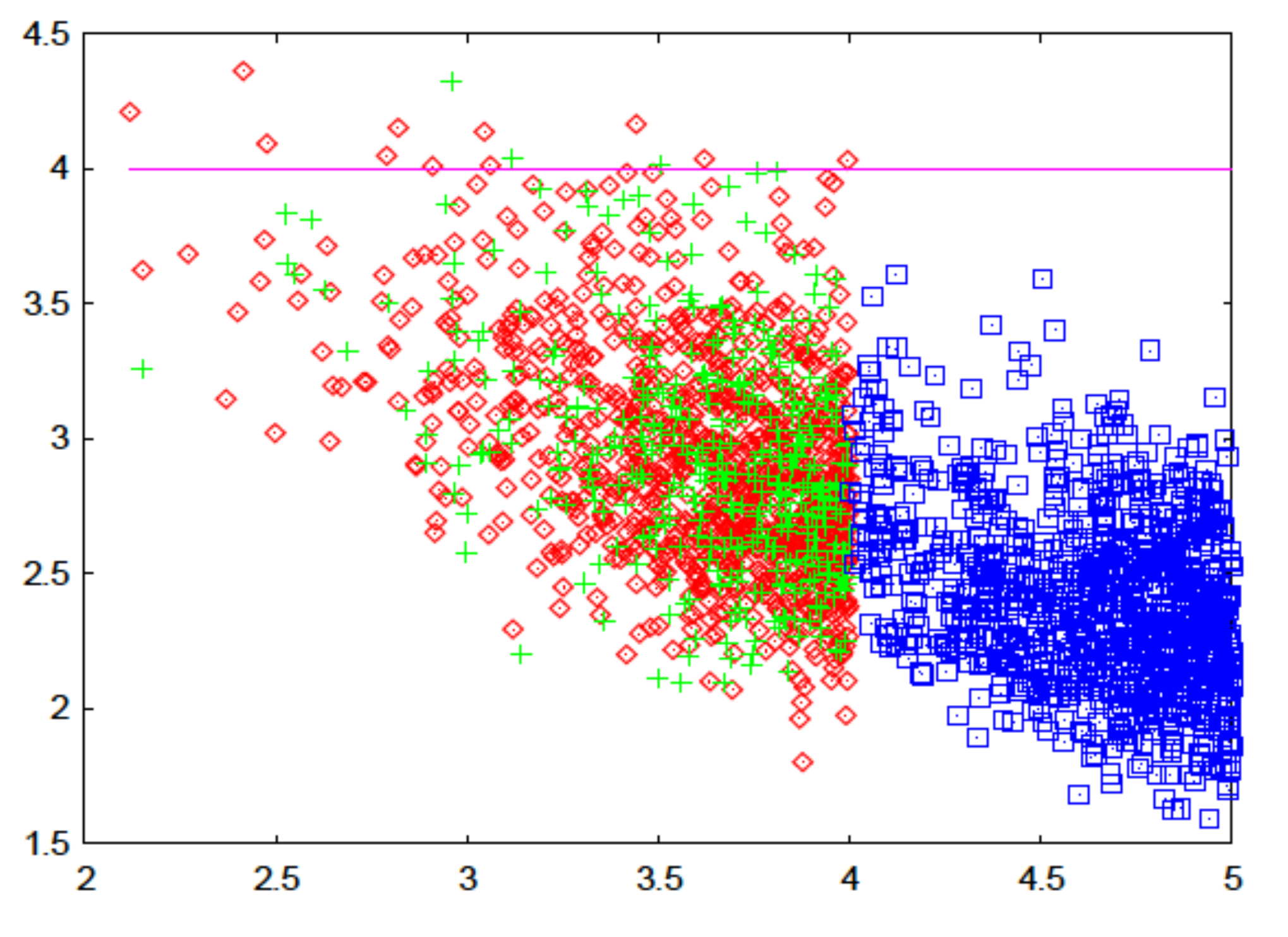}
\caption{(Color online) Plot of the maximum violation of the MABK inequality for $n=4$ qubits vs. $R$. It is apparent that beyond $R=4$ no MABK violation occurs. See text for details.}
\label{fig4}
\end{center}
\end{figure}

\noindent Although the study of arbitrary quantum correlations in higher dimensions becomes really involved, if we take the easy case of generalized Werner states, the maximum 
violation of the concomitant MABK Bell inequalities (let us consider $n$ the number of qubits to be an even number, the well-known ``Ardehali'' MABK Bell
inequality) can be obtained. It is $2^{\frac{n+1}{2}}\cdot p$, linear with the weight $p\in [0,1]$, corresponding to $|GHZ\rangle_n$. Now, when finding the value for the critical value $p_c$ such that below this value no violation occurs (we suppose an idealized case with efficiency of all detectors $\eta=1$), we find that it is exactly the inverse of the amount by which quantum mechanics exceeds LVM predictions, which is an exponentially large factor. Therefore, when considering the number of qubits $n$ large enough, we encounter that $p_c \rightarrow 0$ . That is, as we increase the number of qubits, generalized Werner states tend to have no violation of the concomitant MABK Bell inequalities or, in other words, the range where nonlocality can be found is infinitesimally closer to pure states ($R=1$). This previous fact does not imply that a nonlocal state could be found for $R>1$, but certainly the general trend will be that nonlocality will be found only very close to pure states. And if that is the case, we shall expect both classical entropic inequalities and MABK Bell inequalities to provide the same answer for a given state $\rho$. Summing up, when we are dealing with mixed states, as the number of qubits increases, {\it locality} and {\it classicality} become identical.

\section{Conclusions}

We have explored the whole set of states for different multiqubit systems and have discovered an intimate relationship between locality and classicality. The concomitant results shed new light on the strange connection existing between them when the whole set of states is considered and, in particular, when they are compared with 
states with the same amount of mixedness. The overall overlap of regions where we have states $\rho$ with no entropic violation - no nonlocality - no entanglement as we increase their mixture is quite surprising, although a bit expected for the increasing tendency of states towards maximally mixed ones suggests that no quantum correlations survive. This tendency is checked to extend as we increase the number of parties and, eventually, for higher enough subsystems, they become almost identical. Thus, the non violation of entropic inequalities constitute a close first approximation to detecting quantum correlations and, therefore, reducing the usual number of computations either for Bell correlations or quantum tomography for entanglement-like measurements. 

\section*{Acknowledgements}

J. Batle acknowledges partial support from AUM and the Physics Department, UIB. J. Batle acknowledges fruitful discussions with J. Rossell\'o, Maria del Mar Batle and Regina Batle. Also, J. Batle would like to acknowledge the careful reading of the manuscript done by the referees, as well as their scientific honesty and common sense. 
M. A. would like to acknowledge the financial support from Zewail City for Science and Technology. 
R. O. acknowledges support from High Impact Research MoE Grant UM.C/625/1/HIR/MoE/CHAN/04 from the Ministry of Education Malaysia.

\section*{APPENDIX I: Definition and calculation of quantal measures}

For the quantum
discord QD, a minimization takes place for two parameters only, whereas non-locality
$MABK_{N}^{\max}$ requires an exploration among their corresponding
unit vectors defining the observers' settings. In any case, we have
performed a two-fold search employing i) an amoeba optimization
procedure, where the optimal value is obtained at the risk of
falling into a local minimum and ii) the well known simulated
annealing approach \cite{kirkpatrick83}. The advantage of this
computational 'duplicity' is that we can be confident regarding the
final results reached. Indeed, the second recipe contains a
mechanism that allows for  local searches that eventually can escape
 local optima.

\subsection{Quantum discord}

Quantum discord \cite{olli} constitutes a quantitative measure of
the ``non-classicality" of bipartite correlations as given by the
discrepancy between the quantum counterparts of two classically equivalent
expressions for the mutual information. More precisely, quantum discord is
defined as the difference between two ways of expressing (quantum mechanically)
such an important entropic quantifier. If $S$ stands for the von Neumann entropy,
for a bipartite state $A-B$ of density matrix $\rho$ and reduced
(``marginals") ones $\rho_A$-$\rho_B$, the quantum mutual
information (QMI) $M_q$ reads \cite{olli}

\be \label{uno} M_q(\rho)= S(\rho_A) +  S(\rho_B) - S(\rho), \ee
which is to be compared to its associated classical notion
$M_{class}(\rho)$, that is expressed using conditional entropies.
If a complete projective measurement $\Pi_j^B$ is performed on B
and (i) $p_i$ stands for $Tr_{AB}\,\Pi_i^B\,\rho$ and (ii)
$\rho_{A|\vert \Pi_i^B}$ for $[\Pi_i^B\,\rho\,\Pi_i^B/p_i]$, then
our conditional entropy becomes

\be  \label{unobis}  S(A \vert\,\{ \Pi_j^B \}) =\sum_i\,p_i\,
S(\rho_{A|\vert \Pi_i^B}), \ee so that $M_{class}(\rho)$ adopts
the appearance

\be \label{dos} M_{class}(\rho)_{\{ \Pi_j^B \}} = S(\rho_A)- S(A
\vert\,\{ \Pi_j^B \}). \ee Now, if we minimize over all possible $
\Pi_j^B$ the difference $M_q(\rho)-M_{class}(\rho)_{\{ \Pi_j^B
\}}$ we obtain the quantum discord $\Delta$, that quantifies
non-classical correlations in a quantum system, {\it  including
those not captured by entanglement}. The most general parameterization of the
corresponding local measurements that can be implemented on one
qubit (let us call it B) is of the form $\{ \Pi_B^{0^{\prime}}=I_A
\otimes |0^{\prime}\rangle \langle 0^{\prime}|,
\Pi_B^{1^{\prime}}=I_A \otimes |1^{\prime}\rangle \langle
1^{\prime}|\}$. More specifically we have

\begin{eqnarray} \label{unitarity}
 |0^{\prime}\rangle &\leftarrow& \cos\alpha' |0\rangle + e^{i\beta'}\sin\alpha'|1\rangle \cr
 |1^{\prime}\rangle &\leftarrow& e^{-i\beta'}\sin\alpha'|0\rangle - \cos\alpha' |1\rangle,
\end{eqnarray}
\noindent which is obviously a unitary transformation --rotation in
the Bloch sphere defined by angles $(\alpha',\beta')$-- for the B
basis $\{|0\rangle,|1\rangle\}$ in the range $\alpha' \in [0,\pi]$
and $\beta' \in [0,2\pi)$. The previous computation of the QD has to
be carried out numerically, unless the two qubit states belong to
the class of the so called X-states, where QD is analytic
\cite{Xstates}. One notes then that only
states with zero $\Delta$ may exhibit strictly classical
correlations.

\subsection{Geometric quantum discord}

Despite increasing evidences for relevance of the quantum discord
(Qd) in describing non-classical resources in information
processing tasks, there was until quite recently no
straightforward criterion to verify the presence of discord in a
given quantum state. Since its evaluation involves an optimization
procedure and analytical results are known only in a few cases,
such criteria become clearly desirable. Recently, Datta
advanced a condition for nullity of quantum discord \cite{datta},
and progress was also achieved in \cite{geom} by introducing an
interesting geometric measure of quantum discord (GQD).
Let $\chi$ be a generic $\Delta=0-$state. The GQD measure is then
given by

\be \label{tres} D(\rho)= {\rm Min}_{\chi}[||\rho-\chi||^2],  \ee

\noindent
where the minimum is over the set of zero-discord states $\chi$.
We deal then with  the square of Hilbert-Schmidt norm of Hermitian
operators, $||\rho-\chi||^2= Tr[(\rho-\chi)^2]$ . Dakic et al.
show how to evaluate this quantity for an arbitrary two-qubit
state  \cite{geom,luo}.  Moreover, they demonstrate  the their
geometric distance contains all relevant information associated to
the notion of quantum discord. This was a remarkable feat given
that, despite robust evidence for the pertinence  of the
Qd-notion,
 its evaluation involves optimization
procedures, with analytic results being known only in a few
cases.

Now, given the general form of an arbitrary two-qubits' state
 in the Bloch representation
\ben \label{rhoBloch}& 4\rho=   \mathcal{I} \otimes \mathcal{I} +
\sum_{u=1}^{3} x_u \sigma_u \otimes \mathcal{I} + \sum_{u=1}^{3}
y_u \mathcal{I} \otimes \sigma_i + \cr & +\sum_{u,v=1}^{3} T_{uv}
\sigma_u \otimes \sigma_v, \een \noindent with $x_u=Tr(\rho
(\sigma_u \otimes \mathcal{I}))$, $y_u=Tr(\rho (\mathcal{I}
\otimes \sigma_u))$, and $T_{uv}=Tr(\rho (\sigma_u \otimes
\sigma_v))$, it is found in Ref. \cite{geom} that a necessary and
sufficient criterion for witnessing non-zero quantum discord is
given by the rank of the correlation matrix

\begin{equation} \label{Rmatrix}
\frac{1}{4} \left( \begin{array}{cccc}
1 & y_1 & y_2 & y_3\\
x_1 & T_{11} & T_{12} & T_{13}\\
x_2 & T_{21} & T_{22} & T_{23}\\
x_3 & T_{31} & T_{32} & T_{33}
\end{array} \right),
\end{equation}

\noindent that is, a state $\rho$ of the form (\ref{rhoBloch})
exhibits finite quantum discord iff the matrix (\ref{Rmatrix}) has
a rank greater that two. It is seen that the  geometric measure
(\ref{tres}) is of the final form \cite{geom}

\ben \label{Dfinal} & D(\rho)=\frac{1}{4} \bigg( ||{\bf x}||^2 +
|| T ||^2  - \lambda_{\max} \bigg)=\cr &=
\frac{1}{R}-\frac{1}{4}-\frac{1}{4}\bigg( ||{\bf y}||^2 +
\lambda_{\max} \bigg), \een \noindent where $||{\bf x}||^2=\sum_u
x_u^2$, and $\lambda_{\max}$ is the maximum eigenvalue of the
matrix $(x_1,x_2,x_3)^t (x_1,x_2,x_3) + TT^t$. Here the
superscript $t$ denotes either vector or matrix transposition. The
second expression emphasizes the natural dependence of $D$ on the
participation ratio $R=1/Tr(\rho^2)$. Notice that this measure is
intimately connected with the quantities appearing in
(\ref{Rmatrix}).

\subsection{Violation of MABK inequalities}

Most of our knowledge on Bell inequalities and their quantum mechanical violation is based
on the CHSH inequality \cite{CHSH}. With two dichotomic observables per party, it is the
simplest \cite{Collins} (up to local symmetries) nontrivial Bell inequality for the bipartite case with
binary inputs and outcomes. Let $A_1$ and $A_2$ be two possible measurements on A side whose
outcomes are $a_j\in \lbrace -1,+1\rbrace$, and similarly for the B side. Mathematically, it can
be shown that, following LVM,
$|{\cal B}_{CHSH}^{LVM}(\lambda)|=|a_1b_1+a_1b_2+a_2b_1-a_2b_2|\leq 2$. Since $a_1$($b_1$)
and $a_2$($b_2$) cannot be measured simultaneously, instead one estimates after randomly
chosen measurements the average value ${\cal B}_{CHSH}^{LVM} \equiv \sum_{\lambda} {\cal B}_{CHSH}^{LVM}(\lambda) \mu(\lambda)=
E(A_1,B_1)+E(A_1,B_2)+E(A_2,B_1)-E(A_2,B_2)$, where $E(\cdot)$ represents the expectation value.
Therefore the CHSH inequality reduces to

\begin{equation} \label{CHSH_LVM}
|{\cal B}_{CHSH}^{LVM}| \leq 2.
\end{equation}

Quantum mechanically, since we are dealing with qubits, these observables reduce
to ${\bf A_j}({\bf B_j})={\bf a_j}({\bf b_j}) \cdot {\bf \sigma}$, where ${\bf a_j}({\bf b_j})$
are unit vectors in $\mathbb{R}^3$ and ${\bf \sigma}=(\sigma_x,\sigma_y,\sigma_z)$ are the usual
Pauli matrices. Therefore the quantal prediction for (\ref{CHSH_LVM}) reduces to the expectation
value of the operator ${\cal B}_{CHSH}$

\begin{equation} \label{CHSH_QM}
{\bf A_1}\otimes {\bf B_1} + {\bf A_1}\otimes {\bf B_2}
+ {\bf A_2}\otimes {\bf B_1}  -  {\bf A_2}\otimes {\bf B_2}.
\end{equation}

\noindent Tsirelson showed \cite{Tsirelson} that CHSH inequality (\ref{CHSH_LVM}) is
maximally violated by a multiplicative
factor $\sqrt{2}$ (Tsirelson's bound) on the basis of quantum mechanics. In fact, it is
true that $|Tr(\rho_{AB}{\cal B}_{CHSH})|\leq 2\sqrt{2}$ for all observables ${\bf A_1}$,
${\bf A_2}$, ${\bf B_1}$, ${\bf B_2}$, and all states $\rho_{AB}$. Increasing the
size of Hilbert spaces on either A and B sides would not give any advantage in the
violation of the CHSH inequalities. In general, it is not known how to calculate the best
such bound for an arbitrary Bell inequality, although several techniques have
been developed \cite{Toner}.

A good witness of useful correlations is, in many cases, the violation of a Bell inequality
by a quantum state. Therefore we shall consider the optimization of the violation of the CHSH inequality over
the observer's settings as a definitive measure for both signaling
and quantifying nonlocality in two qubit systems.

We are going to determine which is the maximum expectation value of the CHSH operator (\ref{CHSH_QM}) that a two qubit mixed state
$\rho$ with some degree of mixedness, in this case given by the so called participation ratio $R=1/Tr(\rho^2)$, may have.
Notice that no assumption is needed regarding the state being
diagonal or not in the Bell basis. In order to solve the concomitant variational problem
(and bearing in mind that $B_{CHSH}=Tr(\rho{\cal B}_{CHSH})$), let us first find the state that extremizes
Tr($\rho^2$) under the constraints associated with a given value of $B_{CHSH}$, and the normalization of $\rho$.
This variational problem can be cast as

\begin{equation} \label{var}
\delta \big[  Tr(\rho^2) + \beta Tr(\rho{\cal B}_{CHSH}) - \alpha Tr(\rho)   \big]=0,
\end{equation}

\noindent where $\alpha$ and $\beta$ are appropriate Lagrange multipliers. After some algebra, we arrive at the result

\begin{equation} \label{varResult}
B_{CHSH}^{\max}=\sqrt{Tr[{\cal B}_{CHSH}^2]} \cdot \sqrt{ \frac{4-R}{4R} } = 4\cdot \sqrt{ \frac{4-R}{4R} }.
\end{equation}

\noindent This result is valid for the range $R\in[2,4]$. In the region $R\in[1,2]$ we obtain

\begin{equation} \label{region1}
B_{CHSH}^{\max} = \sqrt{\frac{8}{R}}.
\end{equation} \newline

We shall explore nonlocality in the three qubit case through the violation of the Mermin inequality \cite{Mermin}.
This inequality was conceived originally in order to detect genuine three-party quantum correlations impossible to reproduce
via LVMs. The Mermin inequality reads as $Tr(\rho {\cal B}_{Mermin}) \leq 2$, where ${\cal B}_{Mermin}$ is the Mermin operator

\begin{equation} \label{Mermin}
 {\cal B}_{Mermin}=B_{a_{1}a_{2}a_{3}} - B_{a_{1}b_{2}b_{3}} - B_{b_{1}a_{2}b_{3}} - B_{b_{1}b_{2}a_{3}},
\end{equation}

\noindent with $B_{uvw} \equiv {\bf u} \cdot {\bf \sigma} \otimes {\bf v} \cdot {\bf \sigma} \otimes {\bf w} \cdot {\bf \sigma}$
with ${\bf \sigma}=(\sigma_x,\sigma_y,\sigma_z)$ being the usual Pauli matrices, and ${\bf a_j}$ and ${\bf b_j}$ unit vectors
in $\mathbb{R}^3$. Notice that the Mermin inequality is maximally violated by Greenberger-Horne-Zeilinger (GHZ) states.
As in the bipartite case, we shall define the following quantity

\begin{equation} \label{MerminMax}
 Mermin^{\max} \equiv \max_{\bf{a_j},\bf{b_j}}\,\,Tr (\rho {\cal B}_{Mermin})
\end{equation}

\noindent as a measure for the nonlocality of the state $\rho$. While in the bipartite the CHSH inequality
was the strongest possible one, this is not the case for three qubits. The Mermin inequality is not the only existing Bell inequality for
three qubits, but it constitutes a simple generalization of the CHSH one to the tripartite case. Therefore, it will suffice to use
this particular inequality to illustrate the basic results of the present work.  \newline

The first Bell inequality for four qubits was derived by Mermin, Ardehali, Belinskii and Klyshko \cite{MABK}. It constitutes of
four parties with two dichotomic outcomes each, being maximum for the generalized GHZ state $(|0000\rangle + |1111\rangle)/\sqrt{2}$.
The Mermin-Ardehali-Belinskii-Klyshko (MABK) inequality reads as $Tr(\rho {\cal B}_{MABK}) \leq 4$, where ${\cal B}_{MABK}$ is the MABK operator

\begin{equation} \label{MABK}
\begin{split}
B_{1111}&-B_{1112} -B_{1121}-B_{1211}-B_{2111}-B_{1122}-B_{1212}\\
&-B_{2112}-B_{1221}-B_{2121}-B_{2211}+B_{2222}+B_{2221}\\
&+B_{2212}+B_{2122}+B_{1222},
\end{split}
\end{equation}

\noindent with $B_{uvwx} \equiv {\bf u} \cdot {\bf \sigma} \otimes {\bf v} \cdot {\bf \sigma} \otimes {\bf w} \cdot {\bf \sigma}\otimes {\bf x} \cdot {\bf \sigma}$
with ${\bf \sigma}=(\sigma_x,\sigma_y,\sigma_z)$ being the usual Pauli matrices. As in previous instances, we shall define the following quantity

\begin{equation} \label{MABKMax}
 MABK^{\max} \equiv \max_{\bf{a_j},\bf{b_j}}\,\,Tr (\rho {\cal B}_{MABK})
\end{equation}

\noindent as a measure for the nonlocality content for a given state $\rho$ of four qubits. ${\bf a_j}$ and ${\bf b_j}$ are unit vectors
in $\mathbb{R}^3$. MABK inequalities are such that they constitute extensions of previous inequalities with the requirement that generalized
GHZ states must maximally violate them. New inequalities for four qubits have appeared recently (see Ref. \cite{MABKnew}) that possess some other states
required for optimal violation. In the present study we limit our interest to the MABK inequality, although new ones could be incorporated in
order to offer a broader perspective. However, with respect to entanglement, little is know for the quadripartite case, and thus little
comparison can be done.

The optimization is taken over the two observers' settings $\{{\bf a_j},{\bf b_j}\}$, which are real unit vectors in $\mathbb{R}^3$. We choose them to be
of the form $(\sin\theta_k \cos\phi_k,\sin\theta_k \sin\phi_k,\cos\theta_k)$. With this parameterization, the problem consists in finding
the supremum of $Tr(\rho{\cal B}_{CHSH})$ over the $\{k=1\dotsm 8\}$ angles of $\{  {\bf a_1},{\bf b_1},{\bf a_2},{\bf b_2} \}$
that appear in (\ref{CHSH_QM}).

Optimization of $Mermin^{\max}$ (\ref{MerminMax}) (for states diagonal in the Mermin base of maximally correlated states of three qubits $\rho_{Mermin}^{(diag)}$) is carried out in the same fashion as
in the previous bipartite case. Once the observers' settings $\{{\bf a_j},{\bf b_j}\}$, which are real unit vectors in $\mathbb{R}^3$, are parameterized in spherical coordinates
$(\sin\theta_k \cos\phi_k,\sin\theta_k \sin\phi_k,\cos\theta_k)$, the problem consists in finding the supremum of (\ref{MerminMax}) over the set of $\{k=1\dotsm12\}$
possible angles for $\{  {\bf a_1},{\bf b_1},{\bf a_2},{\bf b_2},{\bf a_3},{\bf b_3} \}$ in (\ref{Mermin}).

\noindent Now, in the case of multiqubit systems, one must instead
use a generalization of the CHSH inequality to N qubits. This is
done in natural fashion by considering an extension of the CHSH or
Mermim inequality to the multipartite case. The first Bell
inequality (BI) for four qubits was derived by Mermin, Ardehali,
Belinskii, and Klyshko \cite{MABK}. One deals with four parties with
two dichotomic outcomes each, the BI being maximum for the
generalized GHZ state $(|0000\rangle + |1111\rangle)/\sqrt{2}$. The
Mermin-Ardehali-Belinskii-Klyshko (MABK) inequalities are of such
nature  that they constitute extensions of older inequalities, with
the requirement that generalized GHZ states must maximally violate
them. To concoct an  extension to the multipartite case, we shall
introduce a recursive relation that will allow for more parties.
This is easily done by considering the operator

\begin{equation}
 B_{N+1}  \propto [(B_1+B_1^{\prime}) \otimes B_N + (B_1-B_1^{\prime}) \otimes B_N^{\prime}] ,
\end{equation}

 \noindent with $B_N$ being the Bell operator for N parties and $B_1={\bf v} \cdot {\bf \sigma}$,
 with ${\bf \sigma}=(\sigma_x,\sigma_y,\sigma_z)$ and ${\bf v}$ a real unit vector. The prime on the operator
 denotes the same expression but with all vectors exchanged. The concomitant maximum value

\begin{equation} \label{MABK_Nmax}
MABK_N^{\max} \equiv \max_{ \bf{a_j},\bf{b_j} }\,\,Tr (\rho {B_{N}})
\end{equation}

\noindent will serve as a measure for the non-locality content of a
given state $\rho$ of N qubits if ${\bf a_j}$ and ${\bf b_j}$ are
unit vectors in $\mathbb{R}^3$. The non-locality measure
(\ref{MABK_Nmax}) will be maximized by generalized GHZ states,
$2^{\frac{N+1}{2}}$ being the corresponding maximum value.

\section*{APPENDIX II. Generation of two-qubits states with a fixed value of the 
participation ratio $R$}

The two-qubits case ($N=2 \times 2$) is the simplest quantum mechanical
system that exhibits the feature of quantum entanglement.
One given aspect is that as we increase the degree
of mixture, as measured by the so called participation ratio
$R=1/$Tr[$\rho^2$], the entanglement diminishes (on average).
As a matter of fact, if the state is mixed enough, that state will have
no entanglement at all. This is fully consistent with the fact
that there exists a special class of mixed states which have maximum
entanglement for a given $R$ \cite{MJWK01}
(the maximum entangled mixed states MEMS).
These states have been reported to be achieved in the
laboratory \cite{MEMSexp} using pairs of entangled photons.
Thus for practical or purely theoretical purposes, it may happen
to be relevant to generate mixed states of two-qubits with
a given participation ratio $R$.

Here we describe a numerical recipe to randomly generate two-qubit states, according
to a definite measure and with a given, fixed value of $R$. Suppose
that the states $\rho$ are generated according to the product measure
$\nu = \mu \times {\cal L}_{N-1}$, where $\mu$ is the Haar measure
on the group of unitary matrices ${\cal U}(N)$ and the standard normalized
Lebesgue measure ${\cal L}_{N-1}$ on ${\cal R}^{N-1}$ provides a reasonable computation of
the simplex of eigenvalues of $\rho$. In this case, the numerical procedure
we are about to explain owes its efficiency
to the following {\it geometrical picture} which is {\it valid only
if the states are supposed to be distributed according to
measure} $\nu$). We shall identify the simplex $\Delta $ with a regular
tetrahedron of side length 1, in ${\cal R}^3$, centered at the origin. Let
  ${\bf r}_i$ stand for the vector positions of the tetrahedron's
  vertices. The tetrahedron is oriented in such a way that the vector
  ${\bf r}_4$ points towards the positive $z$-axis and the vector
  ${\bf r_2}$ is contained in the $(x,z)$-semiplane corresponding to
  positive $x$-values. The positions of the tetrahedron's vertices correspond to
  the vectors

\begin{eqnarray} 
\bf{r_1} &=& (-\frac{1}{2\sqrt{3}},-\frac{1}{2},-\frac{1}{4}\sqrt{\frac{2}{3}}) \nonumber \\
\bf{r_2} &=& (\frac{1}{\sqrt{3}},0,-\frac{1}{4}\sqrt{\frac{2}{3}}) \nonumber \\
\bf{r_3} &=& (-\frac{1}{2\sqrt{3}},\frac{1}{2},-\frac{1}{4}\sqrt{\frac{2}{3}}) \nonumber \\
\bf{r_4} &=& (0,0,\frac{3}{4}\sqrt{\frac{2}{3}}).
\end{eqnarray}

\noindent The mapping connecting the points
  of the simplex $\Delta $ (with coordinates $(\lambda_1,\ldots, \lambda_4)$)
  with the points $\bf r$ within tetrahedron is given by the equations

  \begin{eqnarray} \label{tetra1}
  \lambda_i \, &=& \, 2({\bf r}\cdot {\bf r}_i ) \, + \, \frac{1}{4}
  \,\,\,\, i=1, \dots, 4, \cr
  {\bf r} \, &=& \, \sum_{i=1}^4 \lambda_i {\bf r}_i
  \end{eqnarray}

  \noindent The degree of mixture is characterized by the
  quantity $R^{-1} \equiv Tr(\rho^2) = \sum_i \lambda_i^2$. This
  quantity is related to the distance $r=\mid {\bf r} \mid$
  to the centre of the tetrahedron $T_{\Delta}$ by

  \begin{equation} \label{tetra3}
  r^2 \, = \, -\frac{1}{8} \, + \, \frac{1}{2} \sum_{i=1}^4 \lambda_i^2.
  \end{equation}

\begin{figure}[htbp]
\begin{center}
\includegraphics[width=8.8cm]{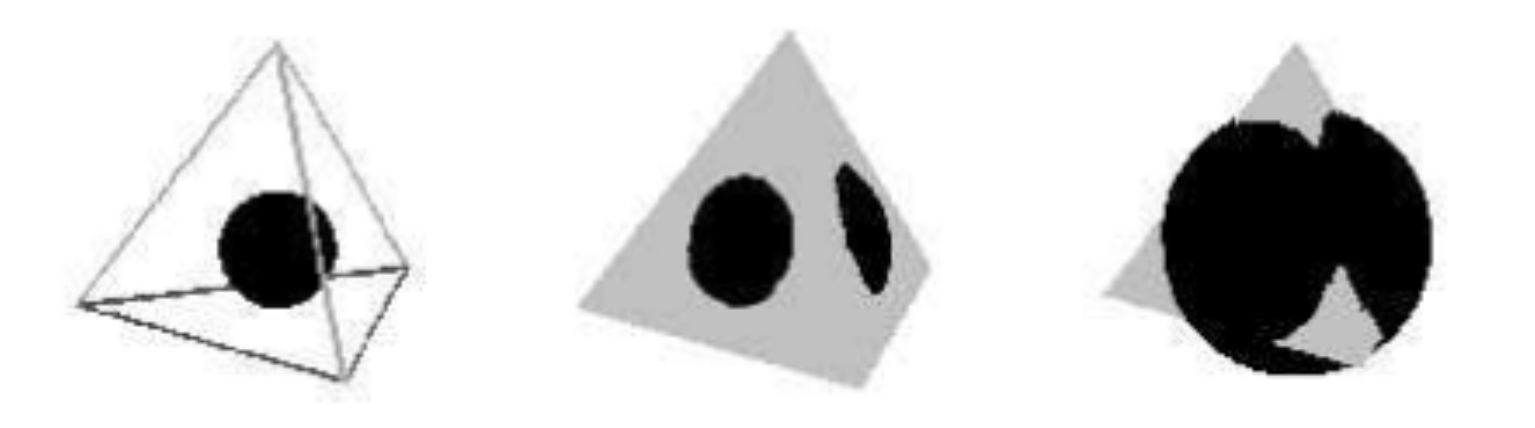}
\caption{(Color online) Geometric picture of the simplex of eigenvalues (tetrahedron) and a sphere whose radius is related to the degree of mixture $R$. Different growing situations correspond to different regions with definite $R$. See text for details.}
\label{tetra}
\end{center}
\end{figure}

 \noindent Thus, the states with a given degree of mixture lie on the
 surface of a sphere $\Sigma_r$ of radius $r$ concentric with the
 tetrahedron $T_{\Delta}$. To choose a given $R$ is tantamount to define
a given radius of the sphere. There exist three different possible regions
(see Fig.\ref{tetra}):

\begin{itemize}

\item region I: $r \in [0, h_1]$ ($R \in [4,3]$), where
$h_1 \equiv h_c={1 \over 4 }\sqrt{2 \over {3}}$ is the radius of a sphere
tangent to the faces of the tetrahedron $T_{\Delta}$. In this case
the sphere $\Sigma_r$ lies completely within the tetrahedron
$T_{\Delta}$. Therefore we only need to generate at random points over its
surface. The Cartesian coordinates for the sphere are given by

\begin{eqnarray} \label{spher}
x_1 &=& r \, \sin\theta\, \cos\phi \nonumber \\
x_2 &=& r \, \sin\theta\, \sin\phi \nonumber \\
x_3 &=& r \, \cos\theta,
\end{eqnarray}

\noindent Denoting {\sf rand\_{}u()} a random number uniformly distributed
between 0 an 1, the random numbers $\phi=2\pi${\sf rand\_{}u()} and
$\theta=\arccos(2${\sf rand\_{}u()}$-1)$ (its probability distribution being
$P(\theta)=\frac{1}{2}\sin(\theta)$) define an arbitrary state $\rho$ on the
surface {\it inside} $T_{\Delta}$. The angle $\theta$ is defined between the
center of the tetrahedron (the origin) and the vector ${\bf r_4}$, and any point
aligned with the origin. Substitution of ${\bf r}=(x_1,x_2,x_3)$
in (\ref{tetra1}) provides us with
the eigenvalues $\{\lambda_i\}$ of $\rho$, with the desired $R$ as prescribed
by the relationship (\ref{tetra3}). With the subsequent application of the
unitary matrices $U$ we obtain a random state $\rho = U D(\Delta) U^{\dag}$
distributed according to the usual measure $\nu = \mu \times {\cal L}_{N-1}$.

\item region II: $r \in [h_1, h_2]$ ($R \in [3,2]$), where
$h_2 \equiv \sqrt{h^{2}_{c}+(\frac{D}{2})^2}={\sqrt{2}\over 4}$ denotes
 the radius of a sphere which is tangent to the sides of the tetrahedron
 $T_{\Delta}$. Contrary to the previous case, part of the surface of the
sphere lies outside the tetrahedron. This fact means that we are able to
still generate the states $\rho$ as before, provided we reject those ones with
negative weights $\lambda_i$.
									
\item region III: $r \in [h_2, h_3]$ ($R \in [2,1]$), where
  $h_3 \equiv \sqrt{h^{2}_{c}+D^2}={\sqrt{6}\over 4}$ is the radius of
a sphere passing through the vertices of $T_{\Delta}$. The generation
of states is a bit more involved in this case. Again
$\phi=2\pi${\sf rand\_{}u()}, but the available
angles $\theta$ now range from $\theta_c(r)$ to $\pi$. It can be shown that
$w\equiv\cos(\theta_c)$ results from solving the equation
$3r^2 w^2 - \sqrt{\frac{3}{2}}r w + \frac{3}{8}-2r^2 = 0$. Thus,
$\theta(r)=\arccos(w(r))$, with $w(r)=\cos\theta_c(r) +
(1-\cos\theta_c(r))${\sf rand\_{}u()}.
Some states may be unacceptable ($\lambda_i<0$) still, but the vast majority
are accepted.

\end{itemize}

\noindent Combining these three previous regions, we are able to generate arbitrary
mixed states $\rho$ endowed with a given participation ratio $R$.

\end{document}